\documentstyle[12pt,epsf]{article}
\begin{document}

\noindent
\begin{large}
{\bf Specificity and stability in topology of protein networks}\\\\
\end{large}

\noindent
{\bf Sergei Maslov}$^1$ \ and {\bf Kim Sneppen}$^2$

\noindent
$^1$ {\sl Department of Physics, Brookhaven National Laboratory, \\
Upton, New York 11973, USA} \\
$^2$ {\sl Department of Physics, Norwegian University of Science
and Technology, N-7491 Trondheim, Norway.}
\vspace{1.0cm}

\centerline{Abstract:}

\noindent
{Molecular networks guide the biochemistry of a living cell
on multiple levels: its metabolic and signalling pathways
are shaped by the network of interacting proteins, whose
production, in turn, is controlled by the
genetic regulatory network. To address topological properties of these
two networks we quantify correlations between connectivities of
interacting nodes and compare them to a null model of a network,
in which al links were randomly rewired.  We find that for both
interaction and
regulatory networks, links between highly connected proteins are
systematically suppressed, while those between a highly-connected
and low-connected pairs of proteins are
favored. This effect decreases the likelihood of cross talk between
different functional modules of the cell, and increases the overall
robustness of a network by localizing effects of deleterious
perturbations.}

\vspace{0.5cm}
With the growth of experimental information about basic biochemical
mechanisms of life, molecular networks operating in living cells are
becoming better defined. Direct physical interactions between
pairs of proteins form one such network.
It serves as a backbone for functional and structural relationships
among its nodes and defines pathways for the propagation of various
signals such as phosphorylation and allosteric regulation of proteins.
The information about specific binding of proteins to each other
has recently grown by an unprecedented amount as
a result of high throughput two-hybrid experiments
\cite{Uetz2000,Ito2001}.
The production and degradation of proteins participating in the
interaction network is controlled by the genetic regulatory network
of the cell formed by all pairs of proteins in which the first
protein directly regulates the abundance
of the second. The majority of known cases of such
regulation happens at the level of transcription, in which a
transcription factor positively or negatively regulates
the RNA transcription of the controlled protein.
The large scale structure of both these networks is
characterized by a high degree of interconnectedness,
where most pairs of nodes are linked to each other by at least one
path. One may wonder how such a heavily intertwined
and mutually dependent dynamical system can perform
multiple functional tasks,
and remain stable against deleterious perturbations.

We analyzed the topological properties of interaction and
transcription regulatory networks in yeast {\it Saccharomyces
cerevisiae}, which at present is perhaps the best
characterized model organism.
The interaction network used in this work consists of
4549 physical interactions between 3278 yeast proteins
as measured in the most comprehensive
two-hybrid screen of yeast proteins \cite{Ito2001}, while the genetic
regulatory network is formed by 1289 directed positive or
negative direct transcriptional regulations within a set of 682 proteins
as listed in the YPD database \cite{Proteome}.
The protein interaction network is a representative of the
broad class of scale-free networks
\cite{Barabasi1999,Broder2000,Jeong2000}
in which the number of nodes with a given number of
neighbors (connectivity) $K$ scales as a power law
$\propto 1/K^{\gamma}$.
In our case the histogram of
connectivities can be fitted by a power law with
$\gamma=2.5 \pm 0.3$ for $K$ ranging from 2 to about 100
\cite{Wagner2001,JMBO2001}.
A small part of the protein interaction network, formed by
proteins known to be localized in the nucleus and to interact with
at least one other nuclear protein, was visualized (Fig. 1).
One striking feature of this graph is the abundance of
highly connected proteins that are mostly connected
to those with low connectivity, and thus well separated from each other.

To test for correlations in connectivities of nodes
for each of the above two networks we calculated the likelihood $P(K_0,K_1)$
that two proteins with connectivities $K_0$ and $K_1$ are connected
to each other by a link and compared it to the same quantity
$P_r(K_0,K_1)$ measured in a randomized
version of the same network. In this ``null model'' network
all proteins have exactly the same connectivity
as in the original one, while the choice of their
interaction partners is totally random. The transcription
regulatory network is naturally directed, while the network of
physical interactions among proteins in principle lacks directionality.
However, for poorly understood
reasons the two-hybrid experimental data have a significant asymmetry
between baits and preys, with bait hybrids being more likely to
be highly connected than their prey counterparts. This can be seen
e.g. in the fact that average connectivity of baits with at least
one interaction partner is close to 3, whereas the same quantity
measured for preys is only  1.8. Since each reported interaction
involves one bait and one prey protein, this asymmetry needs to be
taken into account when constructing an uncorrelated ``null''
model for the interaction network.
For this purpose in our randomization procedure
we would treat the two-hybrid data as a
directed network with an arrow on each edge pointing out
from bait to prey hybrid. Randomized versions of these two
networks were constructed by randomly reshuffling links,
while keeping the in- and out-degree of each node constant.
A convenient numerical algorithm performing such randomization
consists of first randomly selecting
a pair of directed edges A$\rightarrow$B and C$\rightarrow$D
The two edges are then rewired in such a way that
A becomes connected to D, while C to B. However, in case if
one or both of these new links already exist in the network this
step is aborted and a new pair of edges is selected.
This last restriction prevents the appearance of multiple
edges connecting the same pair of nodes.
A repeated application of the above rewiring step leads to a
randomized version of the original network.
Multiple sampling of randomized networks
allowed us to calculate both the average expectation
and the standard deviation for any particular
property of the random network.

Correlations in connectivities manifest themselves
as systematic deviations of the ratio
$P(K_0,K_1)/P_r(K_0,K_1)$ from 1.
We calculated this ratio for
interaction (Fig. 2A) and regulatory (Fig. 2B) networks,
with $K_0$ and $K_1$ being the total number of interaction
partners of two interacting proteins (for the interaction network),
and  out- and in-degrees of two nodes connected by a directed edge
0$ \rightarrow $1 (for the regulatory network).
Thus by the very construction $P(K_0,K_1)$
is symmetric for the physical interaction network but not for
the regulatory network.
We also estimated the statistical significance $Z(K_0,K_1)$
of the above deviations in the interaction (Fig. 2C) and
regulatory (Fig. 2D) networks, by dividing each observed
deviation from the null model by the standard deviation in
multiple realizations of a randomized network.
The combination of these two plots reveals the
regions on the $K_0-K_1$ plane, where connections
between proteins in the real network are
significantly enhanced or
suppressed, compared to the null model.
In particular red regions in the upper left and the lower right corners
reflect the tendency of highly connected nodes (hubs) to associate
with nodes of low connectivity, while the
blue/green region in the upper right corner
reflects the reduced likelihood that two hub centers
are directly linked to each other. One should also note a
prominent feature on the diagonal of the Fig. 2A and 2C
corresponding to an enhanced affinity of proteins with
between 4 and 9 interaction partners to
physically interact with each other.
This feature can be tentatively attributed to the tendency of
members of multi-protein complexes to interact with
other proteins from the same complex. The above
range of connectivities thus correspond to a typical number of direct
interaction partners of a protein in a complex. When we
checked for interactions between proteins in this
range of connectivities we found 39 pairs of
interacting proteins to belong to the same complex in
a recent high throughput study \cite{Gavin2002},
which is 4 times more than
one would expect to find by pure chance alone.

To further quantify  and compare correlation patterns in
interaction and regulatory networks we calculated
the average connectivity
$\langle K_1 \rangle$ of nearest neighbors of a node,
as a function of its own connectivity $K_0$ (Fig. 3A).
In order to simplify the comparison between
two networks here we characterize each
node in the regulatory network by
its total number of neighbors $K=K_{in}+K_{out}$.
For both interaction and regulatory networks
the average connectivity $\langle K_1 \rangle$
shows a gradual decline with $K_0$, which can be fitted with a
power law $\langle K_1 \rangle \propto 1/K_0^{0.6\pm 0.1}$
over approximately two decades.
This observation gives an additional credit to
the affinity between correlation patterns in these two
protein networks visible in Fig. 2.
It was recently found \cite{Vespignani2001} that the
internet, defined as the set of interconnected routers,
in addition to a scale-free distribution of node
connectivities similar to the protein interaction network,
is characterized by the same correlation pattern between
connectivities of neighboring nodes:
$\langle K_1 \rangle \propto 1/K_0^{0.5}$.
This extends by one step an intriguing similarity
in the topology of these networks of completely
different nature.

For the scale-free physical interaction network
we also plotted the probability distribution of the nearest
neighbor connectivity $K_1$, measured separately
for nodes with small connectivity $K_0 \leq 3$, and for
those with large connectivity $K_0 \geq 100$ (Fig. 3B).
In the absence of correlations this conditional probability does
not depend on $K_0$, and is proportional to
$K_1/K_1^{\gamma} \sim 1/K_1^{1.5}$.
This uncorrelated form holds approximately true for neighbors of a
protein with low connectivity.
It is only violated at the far tail of the distribution due to an excess
likelihood of it being connected to a protein with very high connectivity,
as was mentioned above.  On the other hand, the distribution of
connectivities $K_1$ of neighbors of highly connected proteins
scales as $\propto 1/K_1^{2.5}$ and thus differs from that of lowly
connected ones by a factor of $1/K_1$.

When analyzing molecular networks one should consider
possible sources of errors in the underlying data.
Two-hybrid experiments
give rise to false positives of two kinds.
In one case the interaction between proteins is real but it never
happens in the course of the normal life cycle of the cell due to
spatial or temporal separation of participating proteins. In another
case an indirect physical interaction is mediated by one or more
unknown proteins localized in the yeast nucleus.
Reversely, in a high throughput two-hybrid screens one should
expect a sizeable number of false negatives. Primarily a binding
may not be observed if the conformation of the bait or prey
heterodimer blocks relevant interaction sites or if the
corresponding heterodimer altogether fails to fold properly.
Secondly, 391 proteins out of the potential 5671 baits in
\cite{Ito2001} were not tested as possible bait hybrids because
they were found to activate transcription of the reporter gene in
the absence of any prey proteins.

Unlike for the interaction network, our data for the genetic regulatory
network do not come from a single large scale project. Instead, they
represent a collection of numerous experiments performed by different
experimental techniques in different labs.
Therefore, it is
not feasible even to list possible sources of errors
present in such a diverse data set.
In particular one should worry
about a hidden anthropomorphic factor present in such a network:
some proteins just constitute more attractive subjects of research
and are, therefore, relatively better studied than others.
One should also note that the transcription regulation network is
only a subset of a larger genetic regulatory network, which in
addition to transcriptional regulation includes
translational regulation, RNA editing, etc.
An encouraging sign was that when we separately analyzed the
set representing the current knowledge \cite{Proteome} about
this later more complete network, consisting of 1750 genetic
regulations among 848 proteins we reproduced all of our
empirical results for the transcriptional network.

The observed suppression of connections between nearest neighbors of
highly connected proteins is consistent with compartmentalization
and modularity characteristic of control of many
cellular processes \cite{Hartwell1999}.
In fact, it suggests the picture of functional modules of the
cell organized around individual hubs. To further test
the extent of modularity of hubs and their immediate neighborhood
in each network we selected 15 highest connected nodes.
To provide an unbiased sample of hubs from the point of view of
in and out connectivity half of those nodes were selected as
the highest out-degree
hubs (8 baits with $K_{bait} \geq 90$ for the interaction network and 7
nodes with
$K_{out} \geq 34$ for the regulatory network), while half were the highest
in-degree hubs (7 preys with $K_{prey} \geq 20$ for the interaction network
and 8 nodes with $K_{out} \geq 8$ for the regulatory network).
In agreement with the correlation properties described above,
direct connections between
hubs were significantly suppressed. In the interaction network
we observed 20 links between different hubs in this group,
which is significantly  below $56 \pm 7.5$ links
in the randomized network. In the transcription regulatory network
there were 16 links between hubs in real network as opposed to
$35 \pm 6.5$ in its randomized version.
Not only direct links between hubs are suppressed in
both studied networks, but hubs also tend to
share fewer of their
neighbors with other hubs, thereby extending their isolation
to the level of next-nearest neighbor connections.
The total number of paths of length 2 between the
set of 15 hubs in the interaction network is equal to 418,
whereas in the null model we measured this number to be $653\pm 56$.
Similarly, for the transcriptional network
the number of paths of length 2 is equal to
186 in the real network, whereas from the null model one expects
it to be $262 \pm 30$. Since the number of paths of length 2 between
a pair of proteins is equal to the number of their common
interaction partners one concludes that both the hub node itself and its
immediate surroundings tend to separate from other hubs, reinforcing
the picture of functional modules clustered around individual hubs.

A further implication of the observed correlation is in the
suppression of the propagation of deleterious perturbations
over the network. It is reasonable to assume that certain
perturbations such as e.g. significant changes in the
concentration of a given protein (including its vanishing altogether
in a null-mutant cell) with a ceratin probability can affect
its first, second, and sometimes even more distant neighbors in the
corresponding network.  While the number of immediate neighbors
of a node is by definition equal to its own connectivity $K_0$,
the average number of its second neighbors, given
by $K_0 \langle (K_1-1) \rangle_{K_0}$, is sensitive to
correlation patterns of the network. Since highly connected nodes
serve as powerful amplifiers for the propagation of deleterious
perturbations it is especially important to suppress this
propagation beyond their immediate neighbors.
It was argued that scale-free networks in general
are very vulnerable to attacks aimed at highly connected nodes
\cite{Albert2000,Vogelstein2000}. The anticorrelation presented above
implies a
reduced branching ratio around these nodes and thus provides a certain
degree of protection against such attacks. This may be the reason why
the correlation between the connectivity of a given
protein and the lethality of the mutant cell lacking
this protein is not particularly strong \cite{JMBO2001}.

It is feasible that molecular networks in a living cell have organized
themselves in an interaction pattern that is both robust and specific.
Topologically the specificity of different functional modules can be
enhanced
by limiting interactions between hubs and suppressing the average
connectivity of their neighbors.
We have seen that such correlation pattern appears in a similar way in two
different layers of molecular networks in yeast, and thus presumably is
a universal feature of all molecular networks operating in living cells.


\begin{figure}
\epsfxsize=6in
\epsffile{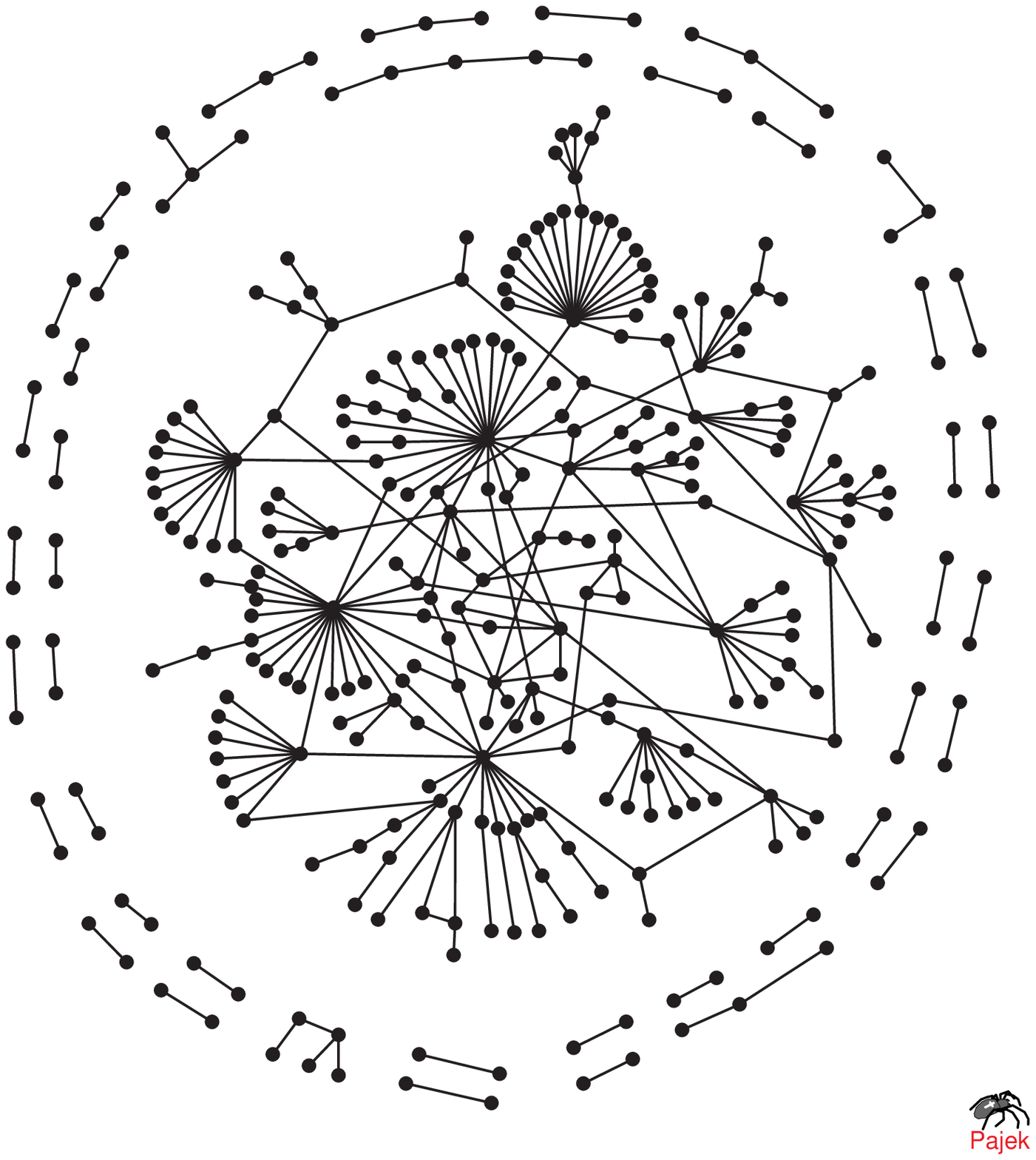}
\caption{Network of physical
interactions between nuclear proteins. Here we show the part of
the network reported in \cite{Ito2001}, consisting of all proteins
that are known to be localized in the yeast nucleus
\cite{Proteome}, and which interact with at least one other
protein in the nucleus. This subset consists of 318 interactions
between 329 proteins.
Note that most neighbors of highly connected nodes
have rather low connectivity.
\label{fig1}}
\end{figure}

\begin{figure}
\epsfxsize=6in
\epsffile{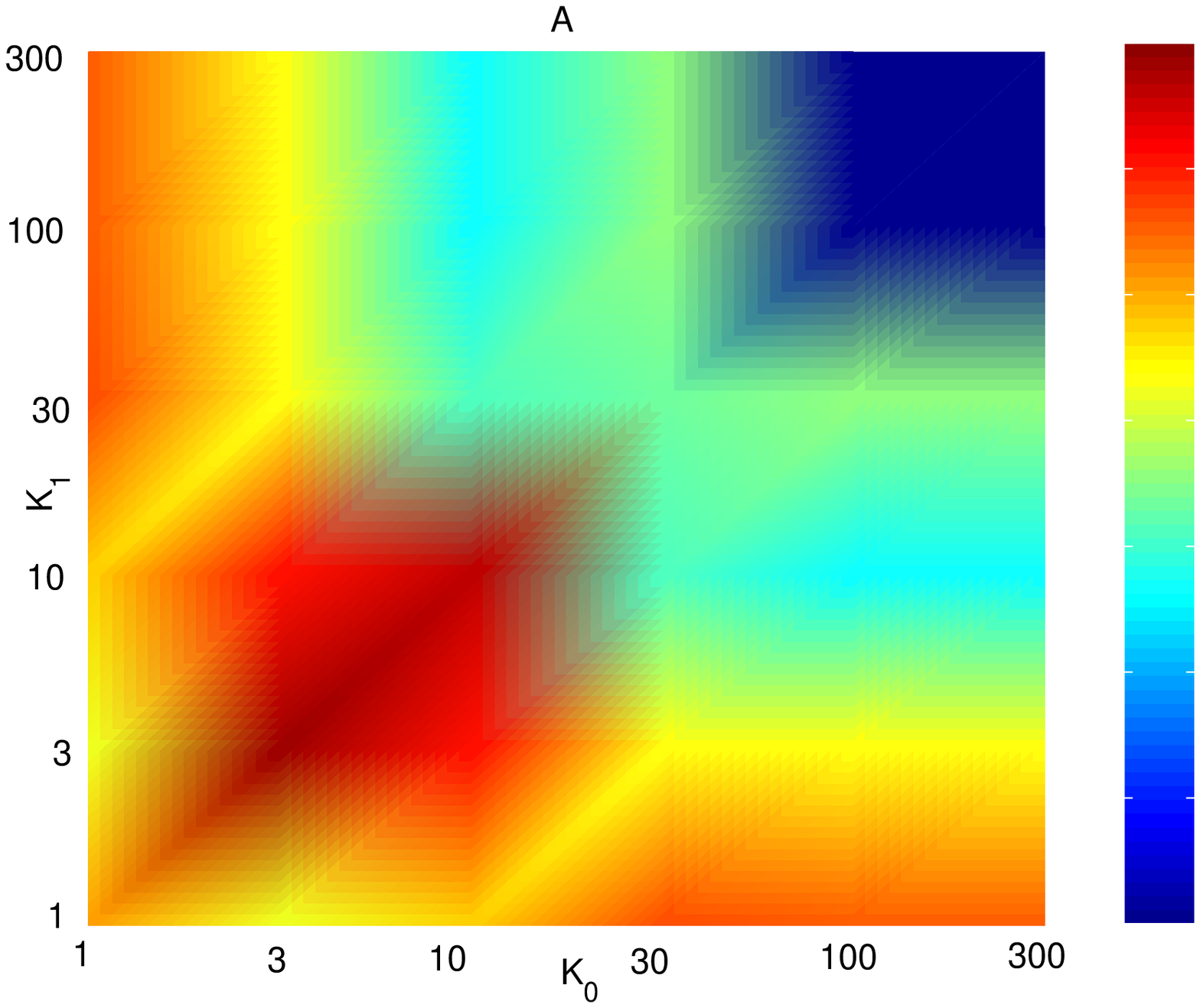}
\end{figure}

\begin{figure}
\epsfxsize=6in
\epsffile{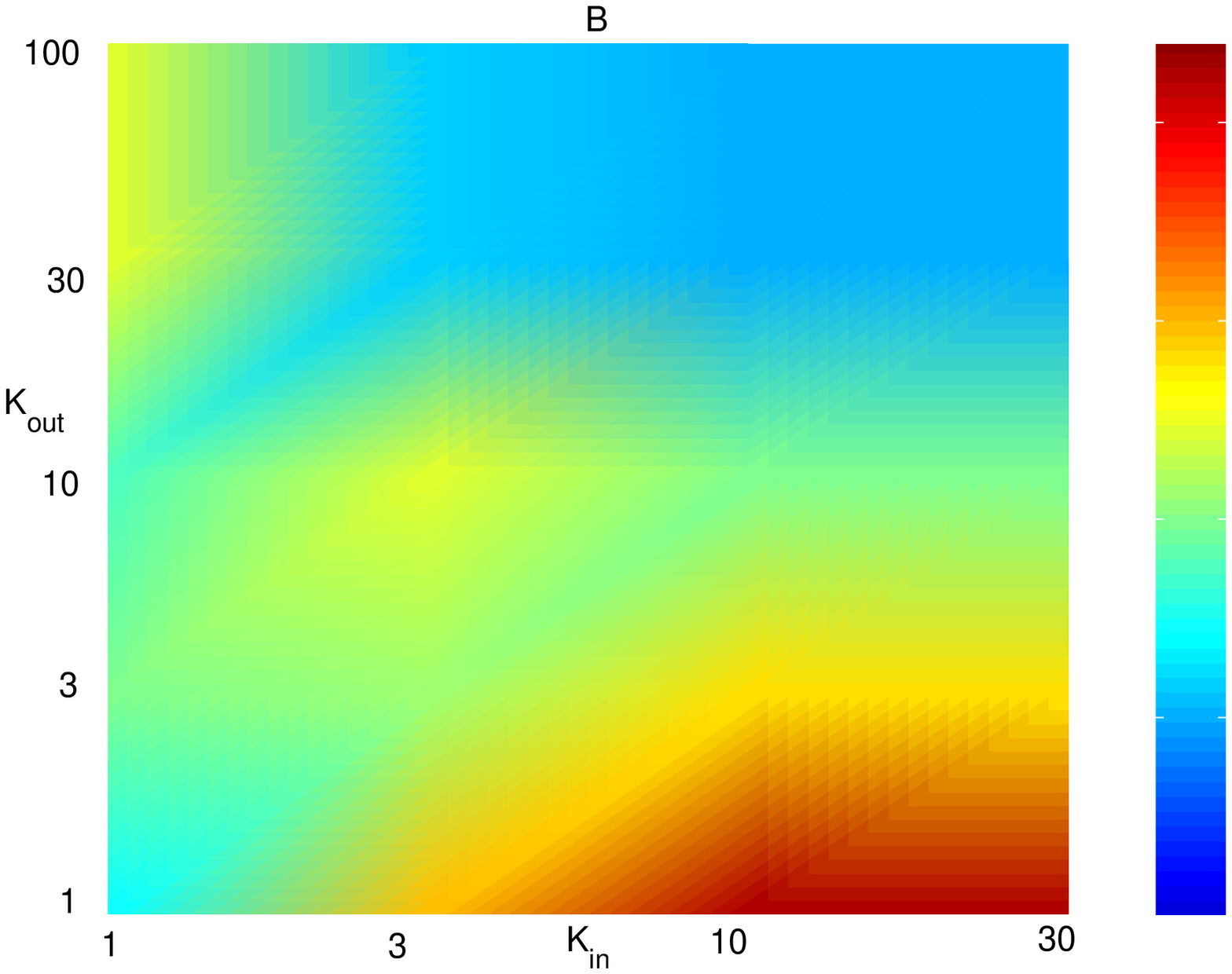}
\end{figure}

\begin{figure}
\epsfxsize=6in
\epsffile{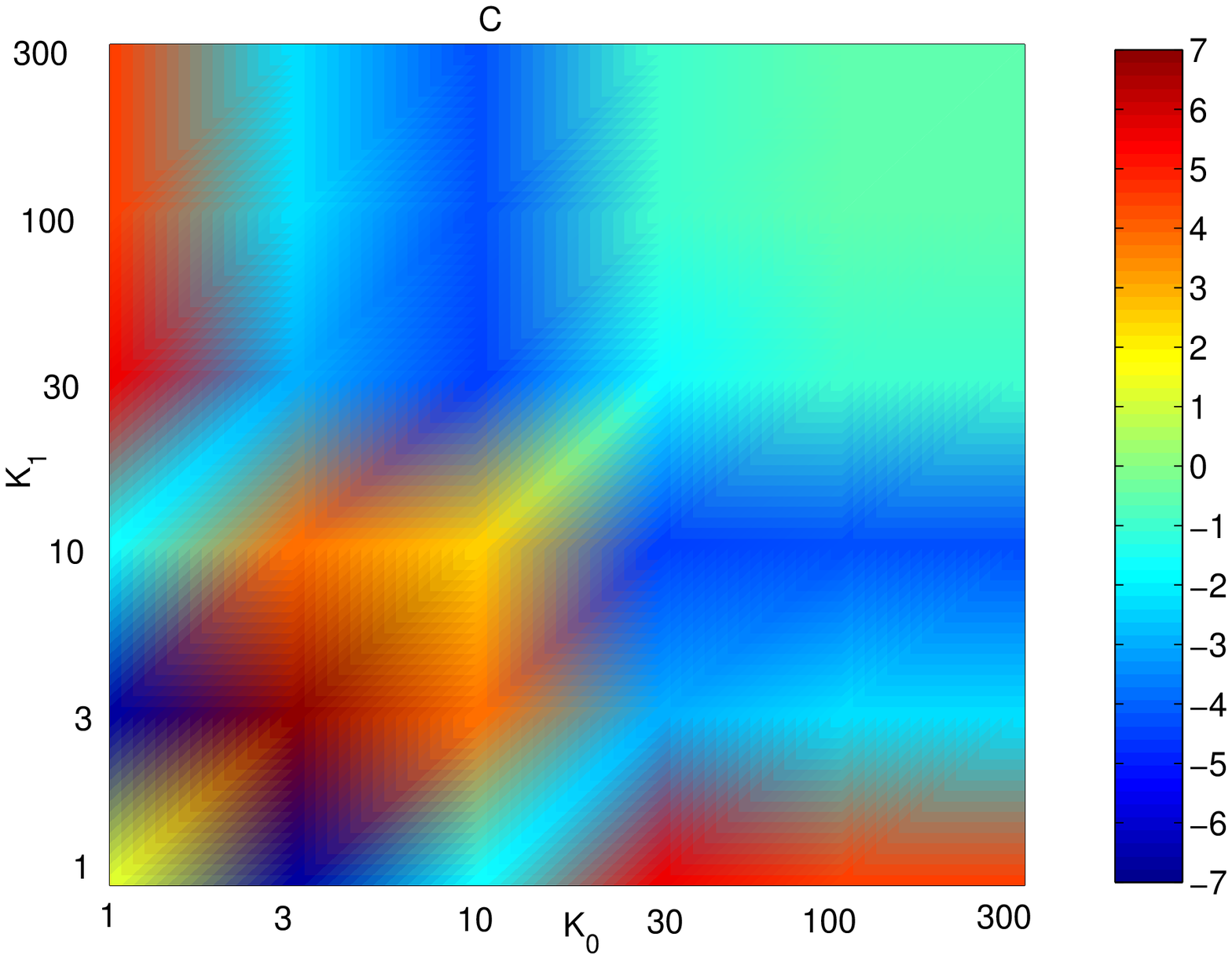}
\end{figure}

\begin{figure}
\epsfxsize=6in
\epsffile{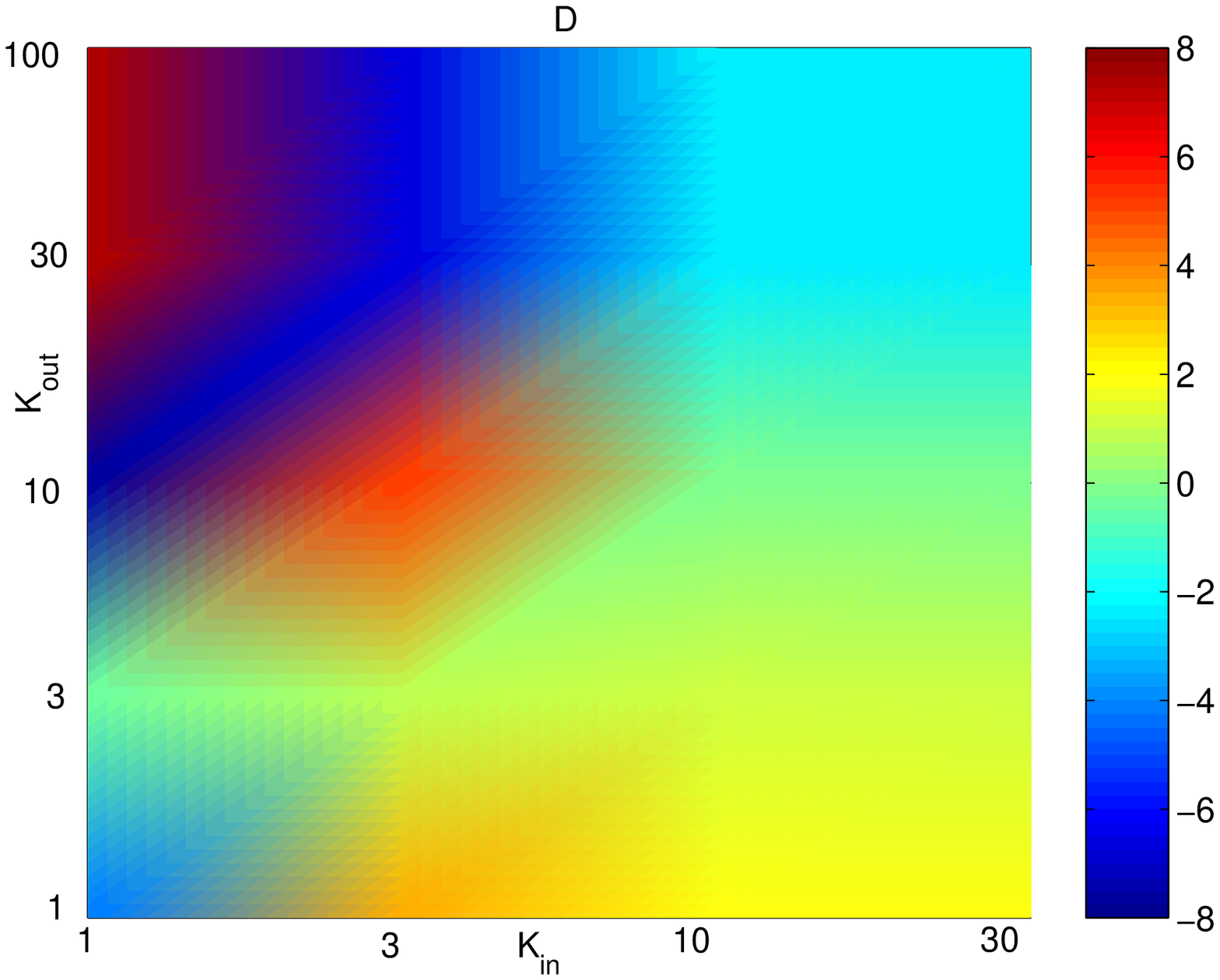}
\end{figure}

\begin{figure}
\caption{
Correlation profiles of protein interaction and regulatory networks
in yeast.
(A) The ratio $P(K_{0},K_{1})/P_r(K_{0},K_{1})$,
where $P(K_{0},K_{1})$ is
the probability that a pair of
proteins with total numbers of interaction partners given by $K_{0},K_{1}$
correspondingly, directly
interact with each other in the full set of \cite{Ito2001},
while $P_r(K_{0},K_{1})$ is the same probability in
a randomized version of the same network.
(B) The same as (A) but for a protein with the in-degree $K_{in}$ to be
regulated by that with the out-degrees $K_{out}$ in
the transcription regulatory network \cite{Proteome}.
(C) Z-scores for connectivity correlations from (A):
$Z(K_{0},K_{1})
=(P(K_{0},K_{1})-P_r(K_{0},K_{1}))/\sigma_r(K_{0},K_{1})$
where $\sigma_r(K_{0},K_{1})$ is the standard deviation of
$P_r ( K_{0}, K_{1} )$ in 1000 realizations of a
randomized network.
(D) As in (C) but for incoming and outgoing links in the
the transcription regulatory network.
To improve statistics the connectivities in all four panels of
Fig. 2 were logarithmically binned into 2 bins per decade.
\label{fig2}
}
\end{figure}

\begin{figure}
\epsfxsize=6in
\epsffile{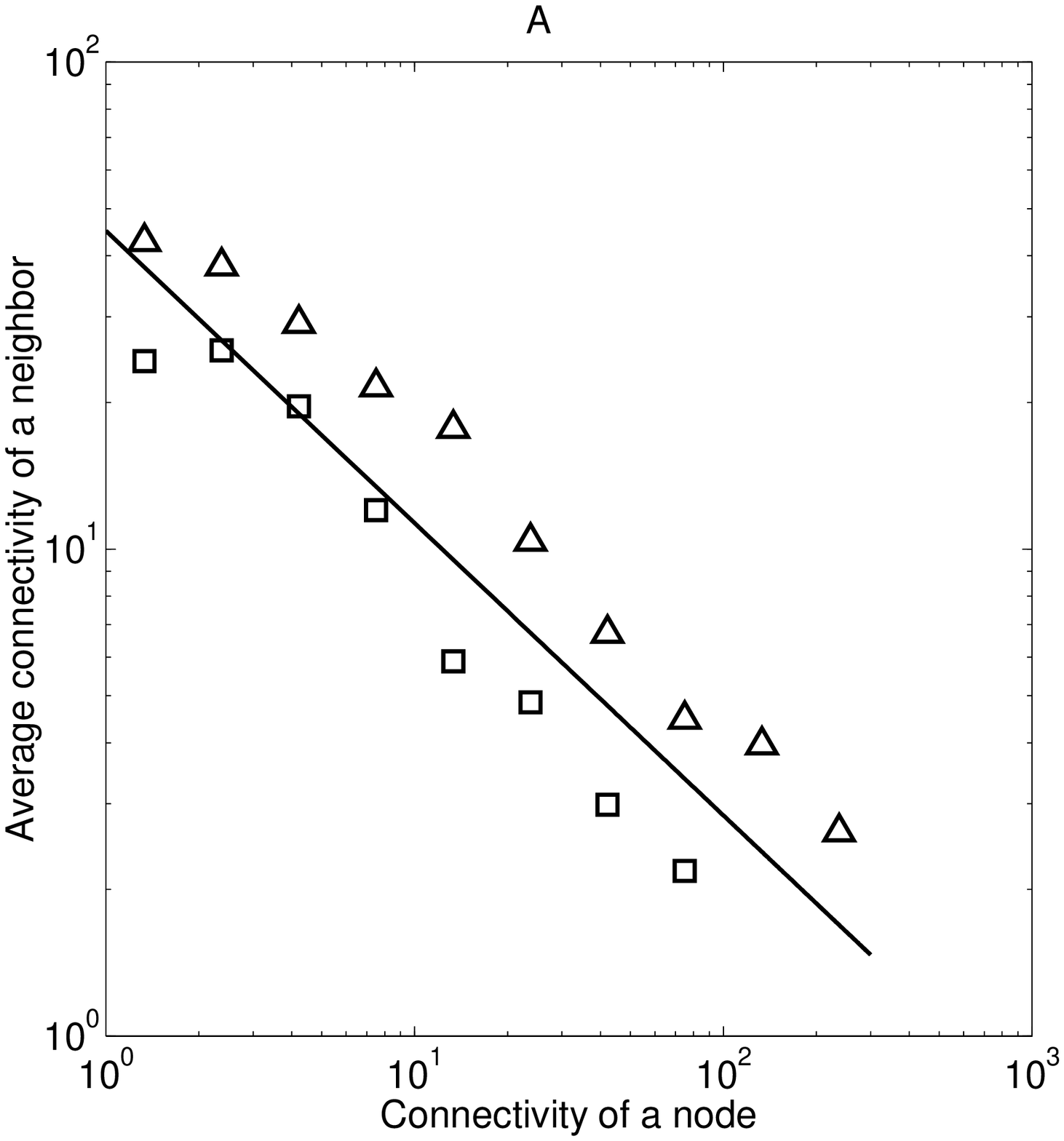}
\end{figure}

\begin{figure}
\epsfxsize=6in
\epsffile{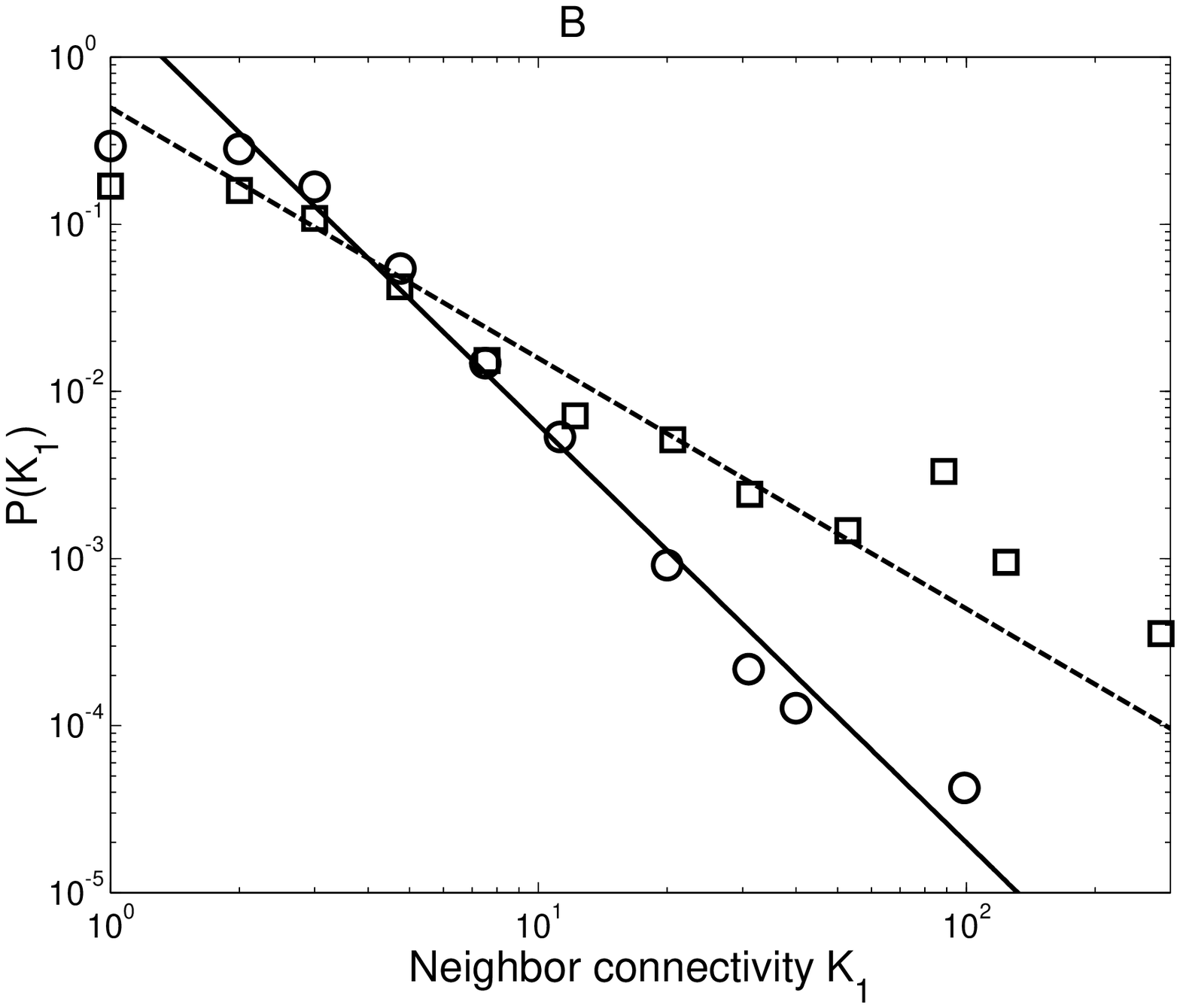}
\end{figure}

\begin{figure}
\caption{Correlations in connectivities of neighbors.
(A) The average connectivity $\langle K_1 \rangle$ of nearest neighbors
of proteins with the connectivity $K_0$ in the physical
interaction network (triangles) and the regulatory
network (squares). The solid line is a power law fit, $\propto 1/K_0^{0.6}$.
(B) The probability distribution of connectivities $K_1$
in the physical interaction network
calculated separately for neighbors of proteins with small
connectivity $K_0\le 3$ (squares), and with large connectivity
$K_0\ge 100$ (circles). Lines are power laws $\propto 1/K_1^{1.5}$ (dashed)
and $\propto 1/K_1^{2.5}$ (solid).
\label{fig3}
}
\end{figure}
\end{document}